\documentclass[12pt,preprint]{aastex}
\usepackage{graphicx}
\usepackage{amsmath}

\def\lapp{\ifmmode\stackrel{<}{_{\sim}}\else$\stackrel{<}{_{\sim}}$\fi}
\def\gapp{\ifmmode\stackrel{>}{_{\sim}}\else$\stackrel{<}{_{\sim}}$\fi}

\begin{document}

\title{A giant glitch in PSR J1718$-$3718} \author{R. N. Manchester and
  G. Hobbs} \affil{CSIRO Astronomy and Space Science, Australia
  Telescope National Facility, Epping NSW 1710, Australia}

\begin{abstract}
  Radio timing observations of the high-magnetic-field pulsar PSR
  J1718$-$3718 have shown that it suffered a large glitch with
  $\Delta\nu_g/\nu = (33.25\pm 0.01)\times 10^{-6}$ between 2007
  September (MJD 54336) and 2009 January (MJD 54855). This is the
  largest pulsar glitch ever observed. As is common, there was a small
  increase in braking torque at the time of the glitch but, unlike all
  other pulsars, the braking torque has continued to increase over the
  two years since the glitch. Polarization observations show that the
  mean pulse profile has about 30\% linear polarization with a smooth
  change of position angle through the pulse and give a rotation
  measure of $-160\pm 22$~rad~m$^{-2}$. There was no detectable change
  in pulse profile at the time of the glitch. The timing observations
  also gave an improved dispersion measure of $371.1\pm
  1.7$~cm$^{-3}$~pc.
\end{abstract}


\section{Introduction}
PSR J1718$-$3718 was discovered in the Parkes multibeam pulsar survey
(PMPS) by \citet{hfs+04}; it is distinguished by its long period, $P
\sim 3.378$~s and very high period derivative, $\dot P\sim 1.598\times
10^{-12}$. Assuming magnetic dipole braking, these parameters imply a
relatively low characteristic age $\tau_c \sim 34,000$~yr and a very
high surface dipole magnetic field $B_s \sim 7.4\times 10^{13}$~G, one
of the highest known for radio pulsars, putting it close to the
magnetars, i.e., anomalous X-ray pulsars (AXPs), and soft gamma-ray
repeaters (SGRs), on the $P - \dot P$ diagram. In fact, X-ray emission
at a similar level to that from quiescent AXPs was detected in {\it
  Chandra} data from a position consistent with that of the pulsar by
\citet{km05}. More recent {\it Chandra} observations have revealed
that this X-ray source is pulsed at PSR J1718$-$3718's period
\citep{zkm+11}, confirming the association. A spectral analysis showed
that the X-ray emission is thermal with a black-body temperature higher
than normal for rotation-powered pulsars of similar age, strengthening the
arguments that PSR J1718$-$3718 is a quiescent magnetar.

Radio timing of this pulsar using the Parkes 64-m radio telescope
commenced soon after the pulsar's discovery in 1999 and continued with
some gaps until 2007 September. In order to provide contemporaneous
timing data for the {\it Chandra} X-ray observations reported by
\citet{zkm+11}, timing observations recommenced in 2009 January. It
was immediately obvious that the pulsar had suffered a massive glitch
since the earlier observations. In this paper we report on the
characteristics of this glitch.\footnote{Given the large gap between
  the last pre-glitch observation and the first post-glitch
  observation, we cannot rule out that there was more than one
  glitch. However we consider this very unlikely as no other glitches
  have been observed over the 12-year data span and large glitches do
  not normally occur in quick succession \citep[see,
  e.g.,][]{elsk11}.} We also give an improved dispersion measure (DM)
for the pulsar and discuss 3.1 GHz (10 cm) pulse polarization results
including the pulsar's rotation measure (RM).

\section{Timing observations and analysis}
Following the pulsar discovery, timing observations were made using
the center beam of the 20-cm Multibeam receiver and the analogue
filterbank (AFB) system used in the PMPS \citep{mlc+01}. The AFB
system has a bandwidth of 288 MHz centered at 1374 MHz and
observations were typically of 10 -- 20 min duration. These
observations ceased in 2007 February. In 2005 December, several timing
observations were made using the Parkes digital filterbank system
PDFB1 with the 10-cm receiver; this receiver has a bandwidth of 1024
MHz centered at 3100 MHz. Several observations of pulsar were also
made with the 20-cm system and PDFB1 between 2007 July and 2007
September. Regular 10-cm timing observations, mainly using PDFB4,
commenced in 2009 January. This system gives 1024 bins across the
profile and 1024 channels across the 1024 MHz bandwidth. The 20-cm
PDFB1 observations were of 10 -- 15 min duration and those at 10 cm of
10 min duration. All PDFB observations were preceded by a 1-min
observation of a pulsed calibration signal injected into the feed
which enables calibration of the system gain and phase response
\citep[see][]{vmjr10}. Observations of Hydra A were used to place the
data on a flux density scale. Data were processed using the {\sc
  psrchive} pulsar analysis system \citep{hvm04} and the timing
analysis was done using {\sc Tempo2} package \citep{hem06,ehm06}. The
Solar-System ephemeris DE405 \citep{sta98b} was used to convert pulse
times of arrival (ToAs) at the Observatory to the Solar-System
barycenter. There were no significant offsets between the ToAs from
the different observing systems. Results are quoted in the barycentric
dynamical time (TDB) system referenced to TT(TAI) and quoted
uncertainties are one standard deviation. In this paper, we discuss
data up to 2011 March 28.

Figure~\ref{fg:nuvar} shows the variations in pulse frequency $\nu = P^{-1}$
and frequency derivative $\dot\nu$ through the observed data span,
obtained from independent fits to short sections of data. The top
panel clearly shows the large step in $\nu$ resulting from the glitch
(the approximate glitch epoch is indicated by the vertical dashed line)
and the middle and lower panels show that $|\dot\nu|$ increased
significantly after the glitch. More interestingly, there appears to
be a significant downward trend in the $\dot\nu$ values after the
glitch, that is, an increase in $|\ddot\nu|$. 

\begin{figure}
\epsscale{1.0}
\plotone{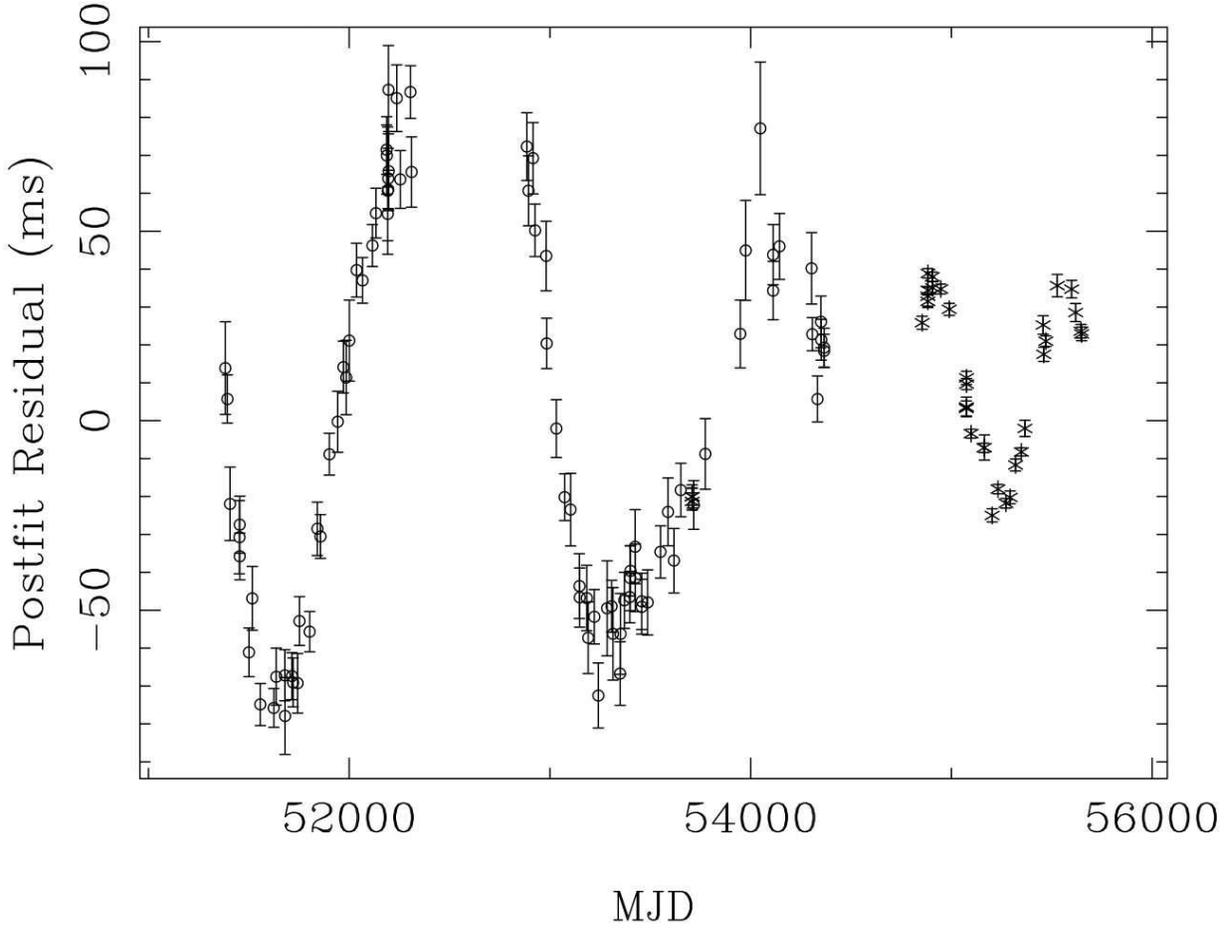}
\caption{Timing residuals for PSR J1718$-$3718 after fitting for pulse frequency
  $\nu$ and its first and second time derivatives and for the jumps in
  these quantities at the time of the glitch using the Cholesky
  method. The glitch occurred in the gap between the higher-quality
  PDFB4 data and the earlier AFB data at about MJD 54600. }\label{fg:res}
\end{figure}

The changes in $\nu$ and its derivatives can be quantified by timing
fits to the pre-glitch and post-glitch data spans; results from such
fits are given in Table~\ref{tb:time}. Uncertainties ($1\sigma$) in
the last quoted digit are given in parentheses. This pulsar exhibits a
substantial amount of timing noise and, for both the pre-glitch and
post-glitch data, the timing parameters were obtained after whitening
of the post-fit residuals by fitting of high-order frequency
derivatives as indicated in Table~\ref{tb:time}. A timing position was
determined from the whitened pre-glitch data: RA(J2000) 17$^{\rm
  h}$~18$^{\rm m}$~09\fs5(6), $-37\degr$~$18\arcmin$~$49\farcs7(23)$.
This position is consistent with but less precise than the {\it
  Chandra} X-ray position reported by \citet{zkm+11}; we therefore
adopt the latter for the subsequent timing analyses. The changes in
the pulse frequency and its derivatives at the time of the glitch
($\Delta\nu_g$, $\Delta\dot\nu_g$ and $\Delta\ddot\nu_g$) can be
estimated from the pre- and post-glitch solutions. It is also possible
to fit directly for them in a timing solution for the whole data
span. Rather than deal with the timing noise by fitting of frequency
derivatives, we have chosen to use the ``Cholesky method''
\citep{chc+11} which is implemented in {\sc Tempo2}. This method
explicity includes a model for the ``red'' timing noise and gives more
realistic estimates of the parameter uncertainties. For PSR
J1718$-$3718, a fit to the power spectrum of the timing residuals
showed that it is well-modelled by a $f^{-3}$ power-law with a cut-off
at 0.2 cycles~yr$^{-1}$ and this was assumed in the Cholesky
fit. Results from this fit are given in the last column of
Table~\ref{tb:time}. A glitch epoch midway between the last pre-glitch
and the first post-glitch ToA was assumed and its quoted uncertainty
covers the gap in the data. Post-fit timing residuals are shown in
Figure~\ref{fg:res}. The resulting values of $\Delta\nu_g$,
$\Delta\dot\nu_g$ and $\Delta\ddot\nu_g$ are consistent with the
differences in $\nu$, $\dot\nu$ and $\ddot\nu$ between the post-glitch
and pre-glitch solutions given in Table~\ref{tb:time} when referred to
the assumed glitch epoch. Figure~\ref{fg:res} shows that the next term
in the Taylor series, $\dddot\nu$, appears to have changed sign from
negative before the glitch to positive afterward. Further observations
are required to verify that this is a glitch-related change.

Comparison of ToAs from 20-cm AFB data and 10 cm PDFB1 data recorded
between 2005 June and 2006 February allowed a more accurate
determination of the pulsar's DM. To compensate for
the effects of interstellar scattering, zero phase of the 20cm
template profile was placed on the leading edge of the profile at
about 60\% of the peak amplitude, leading the peak by 0.005 in
phase. The DM value resulting from the {\sc Tempo2} fit is given in
Table~\ref{tb:par}.

\section{Polarization observations and analysis}
All post-glitch 10-cm PDFB4 observations recorded full Stokes
parameter data. A total of 19 such observations were summed to give a
total integration time of approximately 3.1~h. After calibration for
instrumental gain and phase, the observations were summed in time and
then summed in frequency with a range of rotation measures (RMs) from
$-2000$~rad~m$^{-2}$ to $+2000$~rad~m$^{-2}$ in steps of
40~rad~m$^{-2}$. A clear peak in the linearly polarized intensity
$L=(Q^2 + U^2)^{1/2}$ was observed at about $-155$~rad~m$^{-2}$. Using
this RM, the data were summed to form profiles for two frequency
bands. The weighted mean position angle (PA) difference between the
two bands across the pulse profile and a corresponding $\Delta$RM were
then computed. This process was iterated until convergence to give the
final RM value given in Table~\ref{tb:par}.

\begin{figure}
\epsscale{1.0}
\plotone{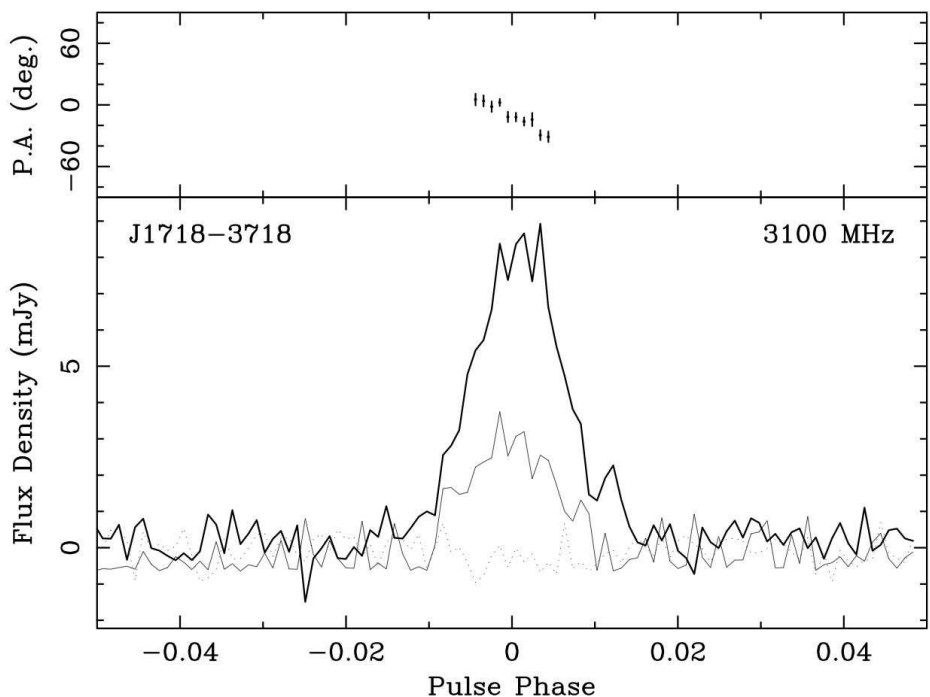}
\caption{Polarization profiles at 10 cm (3100 MHz) for PSR J1718$-$3718. In the
  lower part of the figure the heavier full line is the total
  intensity (Stokes $I$), the lighter full line is the linearly
  polarized intensity ($L = (Q^2+U^2)^{1/2}$) and the dotted line is
  the circularly polarized intensity (Stokes $V$). The upper part of
  the figure shows the position angle of the linearly polarized part
  ($\psi = 0.5 \tan^{-1}(U/Q)$). }\label{fg:poln}
\end{figure}

The final 10-cm polarization profiles shown in Figure~\ref{fg:poln}
were then formed by summing in frequency using the final RM and DM.
Table~\ref{tb:par} gives the mean flux density of the pulsed emission,
the pulse width at 50\% of the peak intensity and the fractional
polarizations, where the $\langle\rangle$ represents a mean value over
the pulse profile. Noise biases in $\langle L\rangle$ and $\langle
|V|\rangle$ have been subtracted \citep[see][]{ymv+11}. There is no
evidence for any change in the 10-cm pulse profile at the time of the
glitch \citep[cf.,][]{wje11} but, because of the low S/N ratio, a
small change cannot be ruled out.

\section{Discussion}
\begin{figure}
\epsscale{1.0}
\plotone{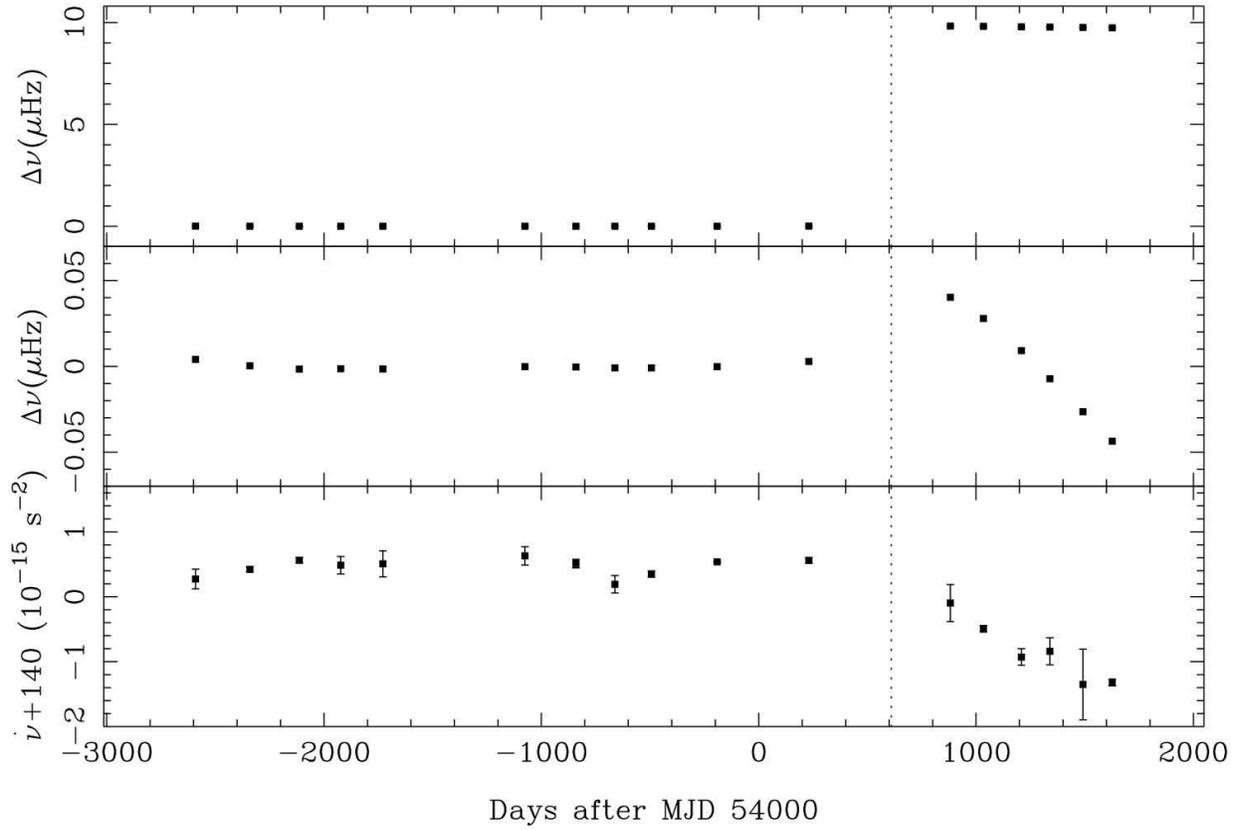}
\caption{Variations of pulse frequency $\nu$ and frequency derivative
  $\dot\nu$ for PSR J1718$-$3718. In the middle plot, the mean post-glitch
  frequency has been subtracted to show more detail. }\label{fg:nuvar}
\end{figure}

The fractional pulse frequency change for this glitch,
$\Delta\nu_g/\nu = (33.25\pm 0.01)\times 10^{-6}$, is the largest ever
observed for any pulsar, magnetars included. The largest previously
known was for PSR B2334+61 which had a glitch with $\Delta\nu_g/\nu
\sim 20.5\times 10^{-6}$ in 2005 \citep{ymw+10} and for magnetars, the
largest glitch so far observed is for 1E~1048.1$-$5937 (PSR
J1048$-$5937) with $\Delta\nu_g/\nu \sim 16.3\times 10^{-6}$
\citep{dkg09}.\footnote{\citet{icd+07} claimed a very large glitch
  ($\Delta\nu_g/\nu \sim 65\times 10^{-6}$) in the AXP CXOU
  J164710.2$-$455216 (PSR J1647$-$4552) but this claim has been
  disputed by \citet{wkga11} who saw no evidence for such a glitch in
  a more complete data set.}  As Figure~\ref{fg:nuvar} and the glitch
parameters in Table~\ref{tb:time} show, there was a step change in
both $\dot\nu$ and $\ddot\nu$ at the time of the glitch. The observed
value of $\Delta\dot\nu_g/\dot\nu \sim (6.0 \pm 0.7)\times 10^{-3}$ is
a little less than values typically observed in other large glitches,
but certainly within the observed range \citep[e.g.,][]{ywml10,elsk11}.

Typically, after a large glitch, the spin-down rate relaxes back
toward the pre-glitch value $\dot\nu_0$ approximately following the relation
\begin{equation}
\dot\nu(t) = \dot\nu_0(t) + \Delta\dot\nu_p + \Delta\dot\nu_d e^{-t/\tau_d}
\end{equation}
where $\Delta\dot\nu_p$ is a quasi-permanent change to $\dot\nu$ at
the time of the glitch and $\Delta\dot\nu_d$ is the amplitude of a
component which decays away on a timescale $\tau_d$ \citep[see,
e.g.,][]{ymw+10,elsk11}. In some cases the $\Delta\dot\nu_p$ is not
really permanent but slowly decays in an approximately linear fashion
\citep[e.g.,][]{sl96,ywml10} and in other cases there is little or no
exponential decay. If present, the exponential term implies an
increase in $\ddot\nu$ at the time of the glitch which decays with the
same time constant. 

The post-glitch behaviour of PSR J1718$-$3718 was very different to
the types of relaxation described above, all of which imply both
$\Delta\ddot\nu_g \ge 0$ and a post-glitch value of $\ddot\nu \ge
0$. For PSR J1718$-$3718 there was a significant {\em decrease} in
$\ddot\nu$ at the time of the glitch and no evidence for any
exponential recovery.  The step decrease in $\ddot\nu$ at the time of
the glitch is shown by the post-glitch downward trend in $\dot\nu$
seen in Figure~\ref{fg:nuvar}, by the difference in the fitted
$\ddot\nu$ values given in Table~\ref{tb:time} for the pre- and
post-glitch data spans and by the value of $\Delta\ddot\nu_g$ from the
glitch fit. It is possible that there was an exponential decay in
$\dot\nu$ that we missed because of the large data gap around the
time of the glitch but, since we see no evidence for it in the
post-glitch data, its decay time constant would have to be $\lapp
250$~d. The change in $\ddot\nu$ has persisted for more than 700 days
and seems quite distinct from the period fluctuations observed prior
to the glitch (Figure~\ref{fg:nuvar}). None of the 32 large glitch
events discussed by \citet{elsk11} show this type of post-glitch
behaviour. 

Glitches in AXPs tend to have unusual properties compared to those in
normal radio pulsars \citep[e.g.,][]{dkg08,lnk+11}. With its relatively long
period, strong implied surface dipole magnetic field and X-ray
emission, PSR J1718$-$3718 has been identified as a possible quiescent
AXP \citep{km05,zkm+11}. The unique post-glitch timing
behaviour of this pulsar may be yet another example of these unusual
properties.

The recoveries observed after most large glitches are attributed to
the changing torque on the neutron-star crust as differential
rotations of the internal superfluids and the crust relax back toward
their equilibrium values \citep[e.g.][]{accp93,ac06}. However in the case
of PSR J1718$-$3718 we do not have a relaxation, we have the opposite
--- the effective braking torque increased by a small amount at the
time of the glitch and has continued to increase since then. The
continued increase implies a negative braking index --- the
post-glitch effective braking index is about $-150$ whereas before the
glitch it was approximately $-15$. The characteristic timescale for
the increase in torque is relatively short: $\dot\nu/\ddot\nu \sim
450$~yr.

A secular increase in the post-glitch braking torque is not predicted
by standard glitch models. In the superfluid model it implies that the
differential rotation of the crust and superfluid interior, decreased
by the glitch, continues to decrease despite the secular slowing of
the crustal rotation rate. The short characteristic timescale for the
post-glitch increase in braking (or equivalently, the large negative
value of the post-glitch braking index) implies that the effect is not
purely electromagnetic. If the glitch resulted in, or was caused by, a
crustal shift as envisaged by \citet{rud91c}, then it is possible that
complex interactions in the interior of the star \citep{rzc98} could
result in an increasing braking torque. Another possibility is that a
heat pulse associated with the glitch could result in an increasing
particle flow in the magnetosphere as the surface temperature
increases. Such thermal flows can have a timescale of years
\citep{ll02}. However \citet{zkm+11} found no evidence for any
increase in the X-ray flux associated with the glitch. Recent studies
of neutron-star superfluidity \citep{pmgo06,asc07,lin11} have
shown that the onset of turbulence in the interior superfluid can
modify the spin behaviour of pulsars following a glitch. Specifically,
the tangled vortices result in an increase in the frictional coupling
between the superfluid and the rest of the star and hence a decrease
in the relative velocity of the two components. If such turbulence
were induced by this large glitch, then it is possible that the
enhanced dissipation could further decrease the differential velocity
resulting in an increasing effective braking torque.

\acknowledgements

We thank our referee for pointing out the possible role of superfluid
turbulence in modifying the post-glitch recovery. RNM is a CSIRO
Fellow based at CSIRO Astronomy and Space Science. GH is supported by
an Australian Research Council QEII Fellowship (project
\#DP0878388). We thank our colleages for help with the Parkes
observations of this pulsar. The Parkes radio telescope is part of the
Australia Telescope, which is funded by the Commonwealth of Australia
for operation as a National Facility managed by the Commonwealth
Scientific and Industrial Research Organisation.


\clearpage

\begin{deluxetable}{lccc}
\tablecaption{Timing parameters for PSR J1718$-$3718\label{tb:time}}
\tablehead{ & \colhead{Pre-glitch} & \colhead{Post-glitch} & \colhead{All data}}
\startdata
R.A. (J2000)\tablenotemark{a} & 17$^{\rm h}$ 18$^{\rm m}$ 09\fs83 & 17$^{\rm h}$ 18$^{\rm m}$ $09\fs83$ & 17$^{\rm h}$ 18$^{\rm m}$ $09\fs83$ \\
Dec. (J2000)\tablenotemark{a} & $-37\degr$ $18\arcmin$ $51\farcs5$ & $-37\degr$ $18\arcmin$ $51\farcs5$ & $-37\degr$ $18\arcmin$ $51\farcs5$ \\
Pulse Frequency $\nu$ (Hz) & 0.29600178047(8) & 0.29598283922(3) & 0.29599421091(13)  \\
Freq. 1st deriv. $\dot\nu$ (10$^{-15}$ Hz$^2$) & $-$139.937(3)  & $-$141.360(4) & $-$139.922(4)  \\
Freq. 2nd deriv. $\ddot\nu$ (10$^{-24}$ Hz$^3$)  & $-$1.1(3)   & $-$9.86(16) & 0.49(6)  \\
Nr of frequency derivatives & 8 & 3 & 2 \\
Epoch of pulse freq. (MJD) & 52874  &  55250 & 53500  \\
Data span (MJD) & 51383 -- 54366   & 54855 -- 55649 & 51383 -- 55649  \\
Rms timing residual (ms) &  8.9  & 2.9  & 43.8  \\
Reduced $\chi^2$/d.o.f.  & 2.3/83         &  4.3/26  & \nodata \\
Glitch epoch (MJD) & \nodata & \nodata & 54610(244) \\
$\Delta\nu_g$ (Hz) & \nodata & \nodata & $9.842(3)\times 10^{-6}$ \\
$\Delta\dot\nu_g$ (10$^{-15}$ Hz$^2$) & \nodata & \nodata & $-0.84(9) $ \\
$\Delta\ddot\nu_g$ (10$^{-24}$ Hz$^3$)& \nodata & \nodata & $-10.6(16)$ \\
\enddata
\tablenotetext{a}{Position from {\it Chandra} X-ray data \citep{zkm+11}}
\end{deluxetable}

\begin{deluxetable}{lc}
\tablecaption{Interstellar and profile parameters for PSR J1718$-$3718\label{tb:par}}
\tablehead{ Parameter & Value}
\startdata
Dispersion measure (cm$^{-3}$ pc) & 371.1(17) \\
Rotation Measure (rad m$^2$)  & $-$160(22) \\
Mean flux density at 3100 MHz $\langle I\rangle$ (mJy) & 0.30 \\
Pulse width at 50\% of peak (ms) & 41 \\
Fractional linear polarization $\langle L\rangle/\langle I\rangle$ (\%) & 29 \\
Fractional circular polarization $\langle V\rangle/\langle I\rangle$ (\%) & 5 \\
Fractional abs. circular polarization $\langle |V|\rangle/\langle I\rangle$ (\%) & 7 \\
\enddata
\end{deluxetable}

\end{document}